\documentclass[]{spie}  

\usepackage{amsmath,amsfonts,amssymb}
\usepackage{graphicx}
\usepackage[colorlinks=true, allcolors=blue]{hyperref}

\def\lsim{\mathrel{\lower2.5pt\vbox{\lineskip=0pt\baselineskip=0pt
          \hbox{$<$}\hbox{$\sim$}}}}

\def\gsim{\mathrel{\lower2.5pt\vbox{\lineskip=0pt\baselineskip=0pt
          \hbox{$>$}\hbox{$\sim$}}}}

\title{Fast resets of sub-windows on infrared detectors as a strategy for persistence mitigation}

\author[a]{Edward L. Chapin$^*$}
\author[b]{Theodore Grosson}
\author[a]{Tim Hardy}
\author[a]{Jordan Lothrop}

\affil[a]{National Research Council Herzberg, 5071 W Saanich Rd, Victoria, V9E 2E7, Canada}
\affil[b]{Department of Physics and Astronomy, University of Victoria, Victoria, V8W 3P2, Canada}

\authorinfo{$^*$ed.chapin@nrc-cnrc.gc.ca, phone +1 250 363 6971}

\begin{document}
\maketitle

\begin{abstract}
Persistence effects in HgCdTe infrared detectors cause significant artifacts that can impact the quality of science observations for up to many hours after exposure to bright/saturating sources. This problem will have a substantially greater impact on viable observing modes for infrared cameras on future ELTs due to the leap in sensitivities that is expected. In this paper we present new results from an updated test system that was previously used to prototype ``on-detector guide windows'' to provide fast T/T feedback to AO systems, interleaved with simultaneous (slow) full-frame readouts for science. We now explore the possibility of continuously resetting these small regions of the detector that are illuminated with a compact source as a strategy for mitigating persistence, using two different detectors. While our results generally show promise for this observing strategy, we found for one of our detectors that the combination of fast localized resets with intense illumination can introduce a potentially problematic persistent change in local reset levels.
\end{abstract}


%
%

\section{INTRODUCTION}
\label{sec:intro}

Persistence is a well-known problem with HgCdTe infrared detectors in which they can exhibit an excess time-varying dark current for up to many hours following illumination by a bright source. This effect is generally understood to be caused by impurities that trap charges during illumination, which is a function of the number of and length of time that these traps are exposed to charges in the photodiodes\cite{smith2008}. In this paper we report on the results of an experiment in which the window mode of a Teleydyne HAWAII-2RG detector is used to perform fast localized resets of the sensor in an illuminated region to determine whether it can be used as a practical mitigation for persistence, in parallel with longer (science) reads of the full chip without resets. This work builds on an existing system for which we first demonstrated this hybrid readout mode to implement ``On-detector Guide Windows'' for the purpose of providing fast tip/tilt feedback to an adaptive optics system using a science detector\cite{chapin2022}. This general concept has been proposed as an observing mode for use in the future InfraRed Imaging Spectrograph (IRIS) for the Thirty Metre Telescope\cite{larkin2020}. Other approaches to persistence mitigation have been explored in the past, such as intentionally illuminating the detector\cite{mcleod2016}. Finally, other authors have experimented with sub-windows on HxRG detectors\cite{loic2005,baril2006,bezawada2006,waterson2006,boss2009,young2012}, including a companion paper in these proceedings which explores the possibility of suppressing sky emission lines to enable longer integrations with ground-based infrared spectrographs\cite{grosson2024}.

\begin{figure}[hbt]
    \begin{center}
        \includegraphics[width=0.8\linewidth]{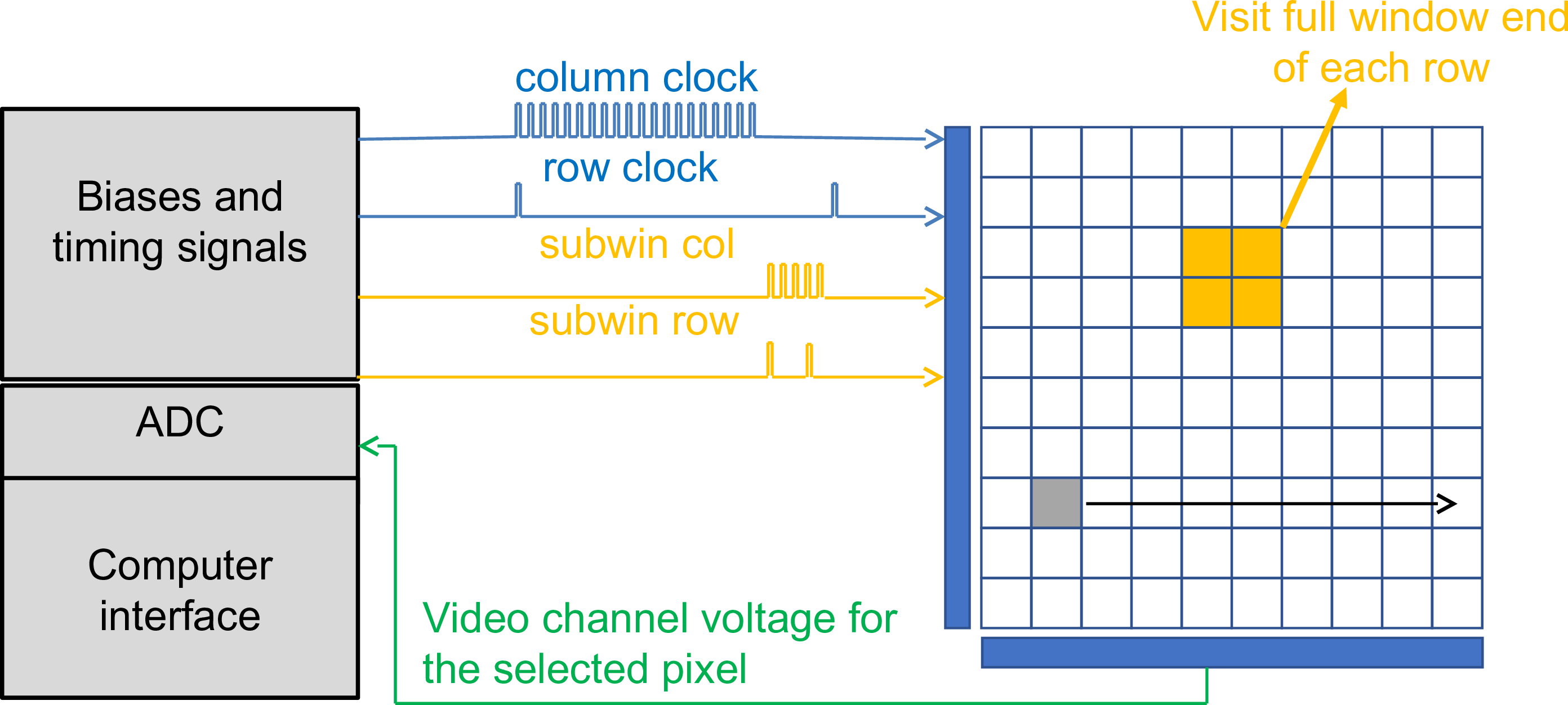}
    \end{center}  
    \caption{\label{fig:readout}
    Illustration of the hybrid Teledyne HAWAII-xRG ``Guide Mode''. The chip has a separate set of registers to address an arbitrary window. Serial commands are issued at the end of each full-frame row to read (and optionally reset) the window before continuing with the next full-frame row. Complete window visits are thus interspersed with readouts of full-frame rows in the output time series.
    }
\end{figure}

To summarize how we control the detector, we implemented the ``Guide Mode'' described in Teledyne documentation for the HAWAII-xRG series of infrared detectors using a Generation 4 (Gen-4) controller from Astronomical Research Cameras. This strategy makes use of the fact that the HAWAII-xRG has separate row and column registers for addressing the full-frame (science field), and a separate (and potentially overlapping) rectangular regions of interest (or windows). Serial commands are issued at the end of each full-frame row to switch into the sub-region mode, at which point the window may be reset and/or read out while the rest of the chip continues to integrate. Figure~\ref{fig:readout} illustrates the main features of this readout strategy. While the ability to perform interleaved non-destructive reads of the window during a science exposure was the main focus of our earlier work (in order to measure time-varying shifts in the centroids of guide stars), in this paper we use the windows to continuously reset regions of the sensor illuminated by a bright source, and then to monitor the dark current in those pixels at a high cadence once the illumination is removed (and the pixels reset). 

In order to perform these more sensitive persistence experiments we made several improvements to our test system. First, we use two significantly better sensors. Both our original detector and the two tested here were purchased by the Canadian Space Agency for the Fine Guidance Sensor (FGS) program for the {\em James Webb Space Telescope (JWST)}. The original detector, and the first of two tested in this work, ``Detector A'', were from early batches that suffered a well-documented indium diffusion problem in their manufacture \cite{rauscher2012}, and we continue to see this problem in our tests. However, the yield of ``Detector A'' compared to the original has increased from roughly 50\% to 75\%. We later switched to a newer detector that was produced after the manufacturing process was improved, and which had been designated as a flight spare for {\em JWST}. We refer to this newer device as ``Detector B'' henceforth, and we only reject a few per cent of its pixels in our analysis. The reason it was not used originally was simply due to the fact that we were missing the flex cable that connects the chip to our cryogenic wiring harness. After failing to procure a replacement, we decided to cannibalize the flex cable from the sensor used in our original experiment. We kept the detectors in their flight packages to minimize development time and the risk of damage. However, a creative approach to mounting them was required to accommodate their attachment flanges which are outward facing, and to include a cover plate (detailed diagrams are shown in our previous paper\cite{chapin2022}). We were able to improve the performance of our $^4$He closed-cycle cryo-cooler and temperature control system to maintain a constant detector temperature of 50\,K for the duration of our experiments. Finally, we mounted an LED in a threaded hole through the cover plate and focused its output using a lens on to the detector to produce a region of compact illumination (for a past experiment that compared localized vs. flatfield persistence measurements see Crouzet et al. 2016\cite{crouzet2016}). For our tests we operate two windows: one located in the centrally illuminated region, and another near the edge of the detector to serve as a reference. While we lacked appropriate flat illumination to infer the gain from photon transfer curves, we adopted a value of 3.3\,e- / ADU as in our earlier experiments, and measurements made with ``Detector B'' from a subsequent experiment are consistent with this value\cite{grosson2024}.

%
%
\section{OBSERVATIONS}
\label{sec:observations}

Prior to each experiment we leave the detector resetting in the dark for a long period of time (at least several hours, and usually over night between tests) to ensure there were no trapped charges before we begin. All of our exposures were then executed using two $8\times8$ windows: one is positioned near the centre of the detector within the region illuminated by the focused LED light, and another is near the edge of the chip where it is dark. Our expectation is that persistence effects should only be detected in the illuminated window, while the dark window should remain unchanged after illumination as a sanity check. We are using a single readout with a 100\,kHz pixel clock. Since we are adding 64 window pixel reads to the end of each full-frame row readout, the time it takes to conduct a single full-frame readout is $[2048 \times (2048 + 64)] \div 100\,\mathrm{kHz} = 43.3\,\mathrm{s}$. The additional time required to issue serial commands to enter/exit window mode at the end of each row is negligible. A single window can be visited 2048 times faster than the full frame, or 47.3\,Hz. Since we are operating two windows (on alternating full-frame rows) they are visited at a rate of 23.6\,Hz each (0.042\,s period).

\begin{figure}[t]
    \begin{center}
        \includegraphics[width=0.99\linewidth]{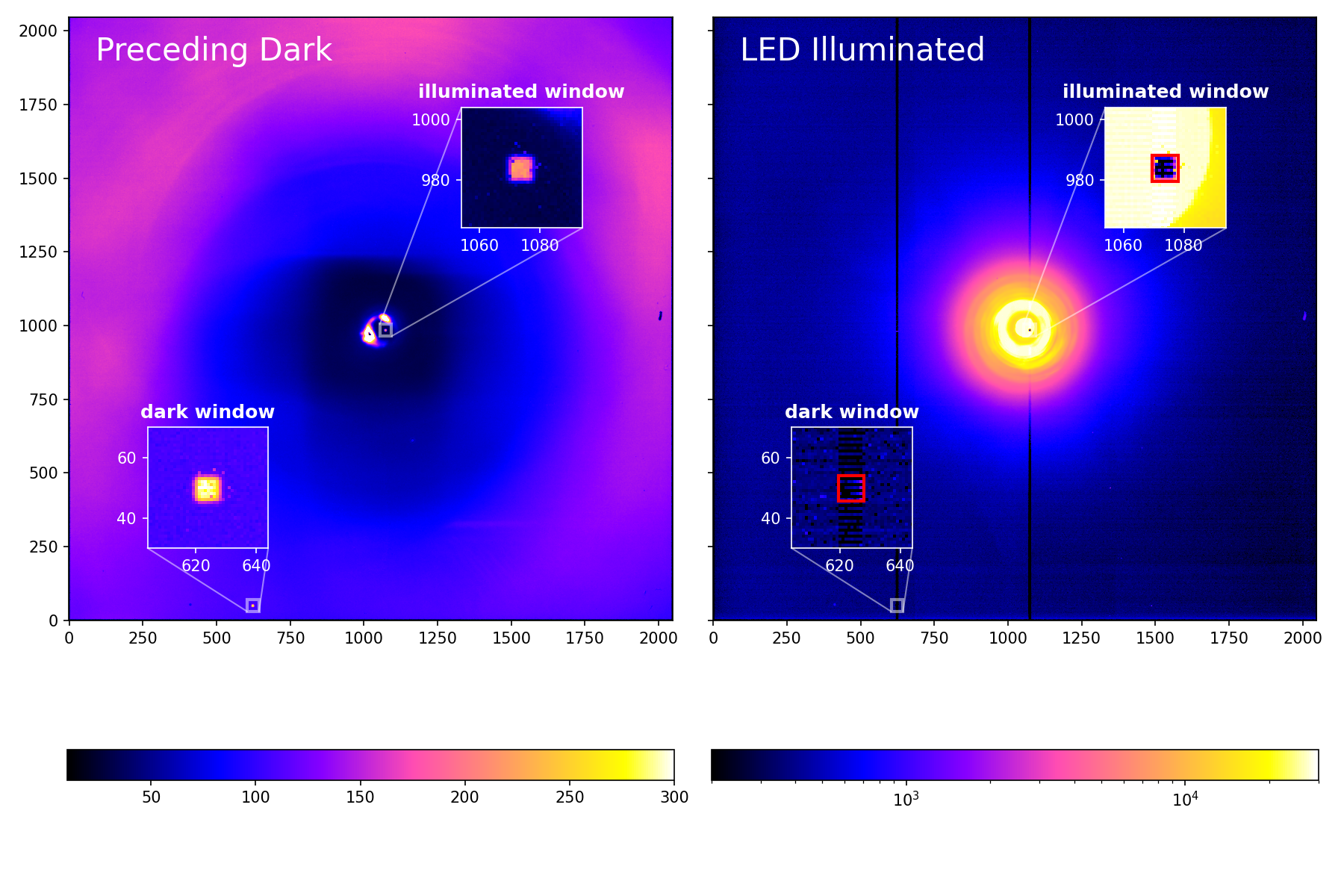}
    \end{center}  
    \caption{\label{fig:overview}
    Example of test observations using Detector B. Left: the dark exposures (linear intensity stretch) exhibit a faint diffuse background with a bright $\sim$100\,pixel diameter ring near the centre that may be caused by thermal conduction from outside the metal cover through the LED wiring. The two $8\times8$ windows are continuously read out which causes notable glow. The ``illuminated window'' is positioned at the centre of the ring so it can provide a useful dark reference prior to activating the LED. Right: for this LED-illuminated exposure (logarithmic intensity stretch) the windows are continuously reset and read out in this example. The vertical dark stripes are reset level artifacts caused by the windows that are corrected in post-processing.
    }
\end{figure}

The observing sequences for our experiments were as follows:

\begin{enumerate}

    \item A 36-minute dark exposure is performed to establish a baseline. For the first full read of the chip, visits to each row are initiated with a line reset followed by reads of every column in that row, and finally reads of the window pixels. There are no further resets of the full frame rows, and no resets are done in the window mode. In total, the resulting data consists of an initial read (immediately after the reset), followed by 50 up-the-ramp (UTR) samples for the full frame. There are 5224 window reads during this time, though up to the first 1024 reads are discarded as the line-by-line (full frame) reset will not yet have reached them.
    
    \item The detector is fully reset, and then illuminated for either 3 or 24 full-frame reads (135 or 1080\,s, respectively), depending on the experiment. During this time the windows are operated in one of the following ways: (i) read every visit; (ii) reset and read every visit (0.042\,s reset interval); (iii) reset every 16 visits  (0.672\,s interval) and read every visit; or (iv) reset every 128 visits (5.376\,s reset interval) and read every visit .
    
    \item The LED is deactivated, and a following 36-minute dark exposure (with an initial reset) is executed, as in Step 1.
    
\end{enumerate}

The illumination times in Step 2 are varied for different experiments to implement different ``soak'' times, the period during which traps are exposed to photoelectrons between resets. Tulloch (2018)\cite{tulloch2018} previously showed that the magnitude and duration of persistence is highly correlated with the soak time. The reason we conduct experiments with different window reset rates is to get a sense of how important the integrated signal is on the observed persistence (i.e., is persistence caused only when pixels reach saturation). Fast resets ensure that there is a very low average signal, while slower resets can result in the pixels attaining a significant fraction of, or even reaching saturation for brief periods. The lower the reset rate that is required, the more time that is available to spend visiting greater numbers and/or larger windows as part of the readout sequence. Finally, we always read the windows on every visit so that we can best sample the behavior of the pixels on short timescales.

Figure~\ref{fig:overview} shows an example (preceding) dark and LED exposure (with window continuously reset) for Detector B. The darks show a faint diffuse pattern of background emission that decreases toward the centre of the detector, but with a prominent ring of bright emission in the middle. We confirmed that this ring was always present when the LED was off, even when physically disconnected from the power supply. It may be caused by thermal emission from the LED caused by conduction from outside the LED and cover assembly through its wiring. It was therefore important to position the ``illuminated'' window at the centre of this ring where it is dark since the bright portions of the ring would actually saturate within the dark exposure time. The ``dark'' window is positioned near the edge of the chip. During the dark integration we continuously read-out the windows which causes obvious glow (a well-known effect\cite{smith2012,regan2020,chapin2022}). The intensity of the dark image is the flux inferred from full polynomial fits to the 50 up-the-ramp (UTR) reads as described in Section~\ref{sec:analysis}.

The LED exposure in Figure~\ref{fig:overview} has a much greater contrast and is displayed with a logarithmic stretch. Since the windows are continuously reset in this example, they appear as dark ``holes'' in the full-frame (science) image. The brightness of the LED was adjusted by placing resistors in series with its power supply, and the peak saturates in $\sim$3\, s (approximately 7\% of the full-frame read time). Since the central region saturates so quickly, we simply show the correlated double-sampled image (CDS, difference of the first two reads after the reset) for the LED exposure, minus the CDS image from the preceding dark. This minimal data reduction clearly leaves strong vertical stripes that are coincident with the window columns. These and other artifacts that can be seen at fainter levels (such as alternating horizontal stripes in the zoomed-in window insets) are reset level artifacts that are effectively removed through the ramp fitting, as evidenced by the dark image on the left in which these artifacts are absent, and is consistent with our previous work\cite{chapin2022}.

\begin{figure}[hbt]
    \begin{center}
        \includegraphics[width=0.9\linewidth]{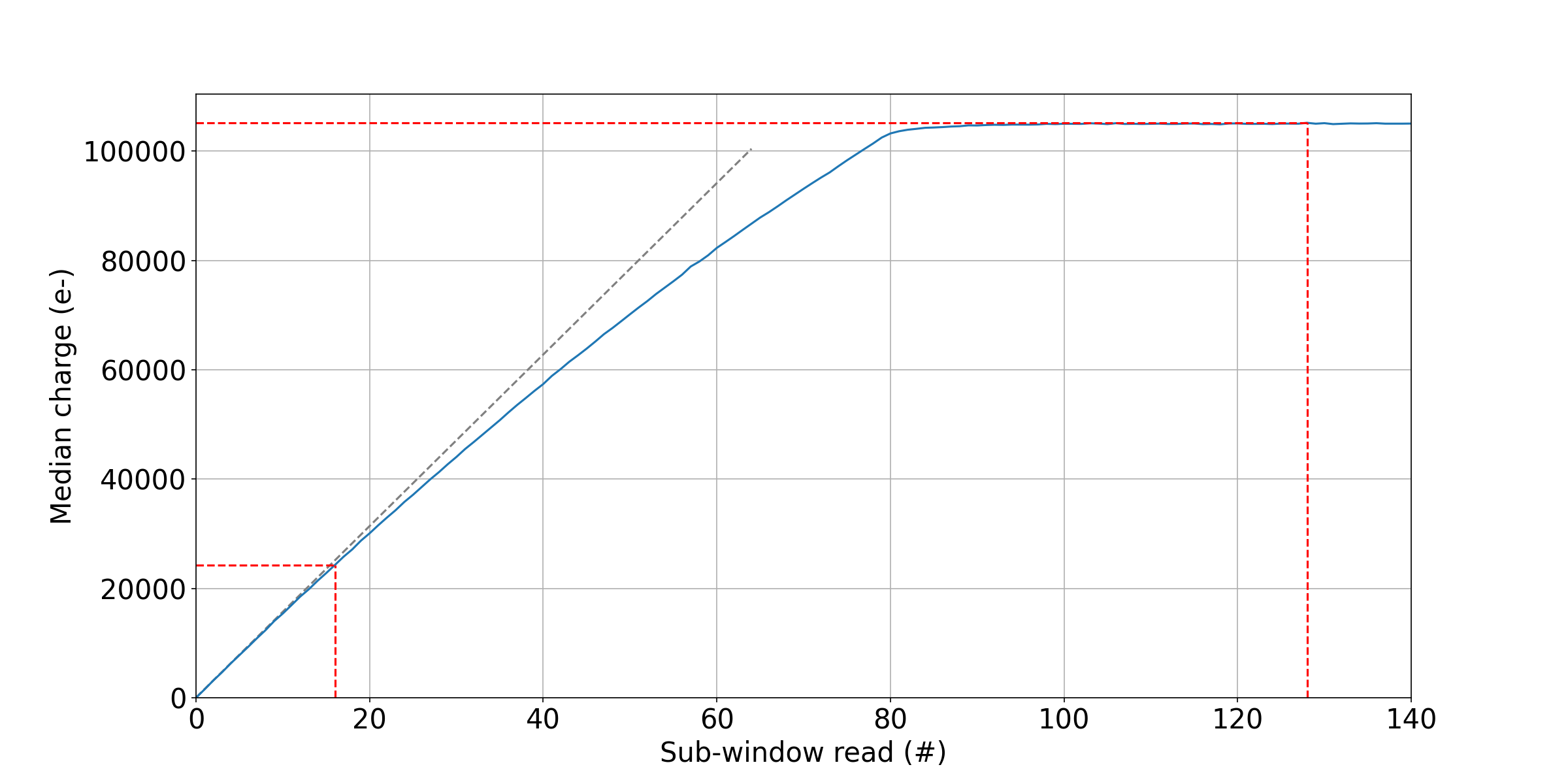}
    \end{center}  
    \caption{\label{fig:ledWinRamp}
    Median UTR samples for the 64 window pixels centered over the illuminated region (uncorrected for non-linearity) when the LED is on. The full-well depth is about 105000\,e-. The red dashed lines indicate reset intervals of 16 and 128 visits which were used for some of our experiments. The grey dashed line shows the linear portion of a 2nd-order polynomial fit to the first 60 reads (the full fit is not shown as it is nearly indistinguishable over that range in this plot).
    }
\end{figure}

The different window reset rates in each experiment limit the peak signal in the illuminated window pixels when the LED is on. Figure~\ref{fig:ledWinRamp} shows the median UTR curve combining all 64 pixels in the illuminated window (after subtracting the initial read as an estimate of the reset level). In this example the windows were not reset, and thus fully sample the ramp from reset to saturation. It shows that the full-well depth is about 105000\,e-, and saturation is reached after about 80 reads (1.92\,s). The linear portion of a 2nd-order polynomial fit is shown as a grey dashed line to give a sense of the non-linear response. Note that the median signal after a single read (what would be expected when the windows are reset at each visit) is 1550\, e- (1.5\% of the full-well depth). After 16 reads it is 22800 (22\% of the full-well depth), and after 128 reads it will have already been saturated for $\sim$ 50 reads.


%
%
\section{ANALYSIS}
\label{sec:analysis}

The following data reduction steps were performed for each observing sequence. First, low-frequency (correlated) electrical readout noise was removed by subtracting the mean value of the reference pixels on each row (4 columns at each end). Some residual horizontal striping is still apparent, but this procedure removes the vast majority of the patchiness that could be seen in residual maps once ramp models were fit. Note that these reference values are removed both from the full-frame row reads, as well as the following 64-pixel window reads at the row ends.

Next, 2nd-order polynomials were fit to each of the full-frame ramps in the initial (pre-illumination) dark exposures (reset and immediate read, followed by 50 UTR reads),
\begin{equation}
\label{eq:ramp}
f(x) = a_0 + a_1x + a_2x^2 ,
\end{equation}
where $x$ is the read number, the reset level is inferred from $a_0$, the flux from $a_1$, and the quadratic correction term from $a_2$ (which primarily accounts for the effect of the shrinking depletion region as the photodiode accumulates electrons). We assume equal weighting in the fits which is sufficient for our purposes, though correlations between reads could be accounted for to improve the result\cite{robberto2009,grosson2024}. For each ramp we perform an iterative fit with 3-$\sigma$ outlier clipping to remove spikes. We also build up an overall bad pixel mask by rejecting pixels with poor $\chi^2$, or outlier flux or reset level values from the fits. The initial read (very shortly after the reset) is affected by the reset anomaly, but given the number of samples in our ramps it has a negligible impact on the fits. For Detector A about 25\% of the pixels are flagged as bad with this procedure, while only 4\% are flagged for Detector B.

In order to measure differences between the darks that precede and follow LED illumination we found that it was necessary to account for the non-linearity in the pixel ramps. This was for two reasons: (i) there were sometimes large differences in the reset levels (higher reset levels require greater non-linearity corrections); and (ii) detrapped persistence charge would cause the following dark pixels to move up the ramp faster, again, further into the non-linear regime. To linearize the data we first fit Equation~\ref{eq:ramp} to the preceding darks, during which we assume the pixels are exposed to a constant flux (due to background light, and also self-heating from the window reads). We then created a lookup table (LUT) to convert from from $f(x)$ to $f'(x) = a_0 + a_1x$ (i.e., the linear portion of the fit), and used the same per-pixel LUT to calculate linearized values for both the preceding and following dark UTR samples. The reset level for each set of corrected UTR samples is estimated from the median of the first 10 reads and removed. Finally, the difference between the preceding and following ramps gives the time evolution of detrapped charge post-illumination for each pixel.

\begin{figure}[hbt]
    \begin{center}
        \includegraphics[width=0.8\linewidth]{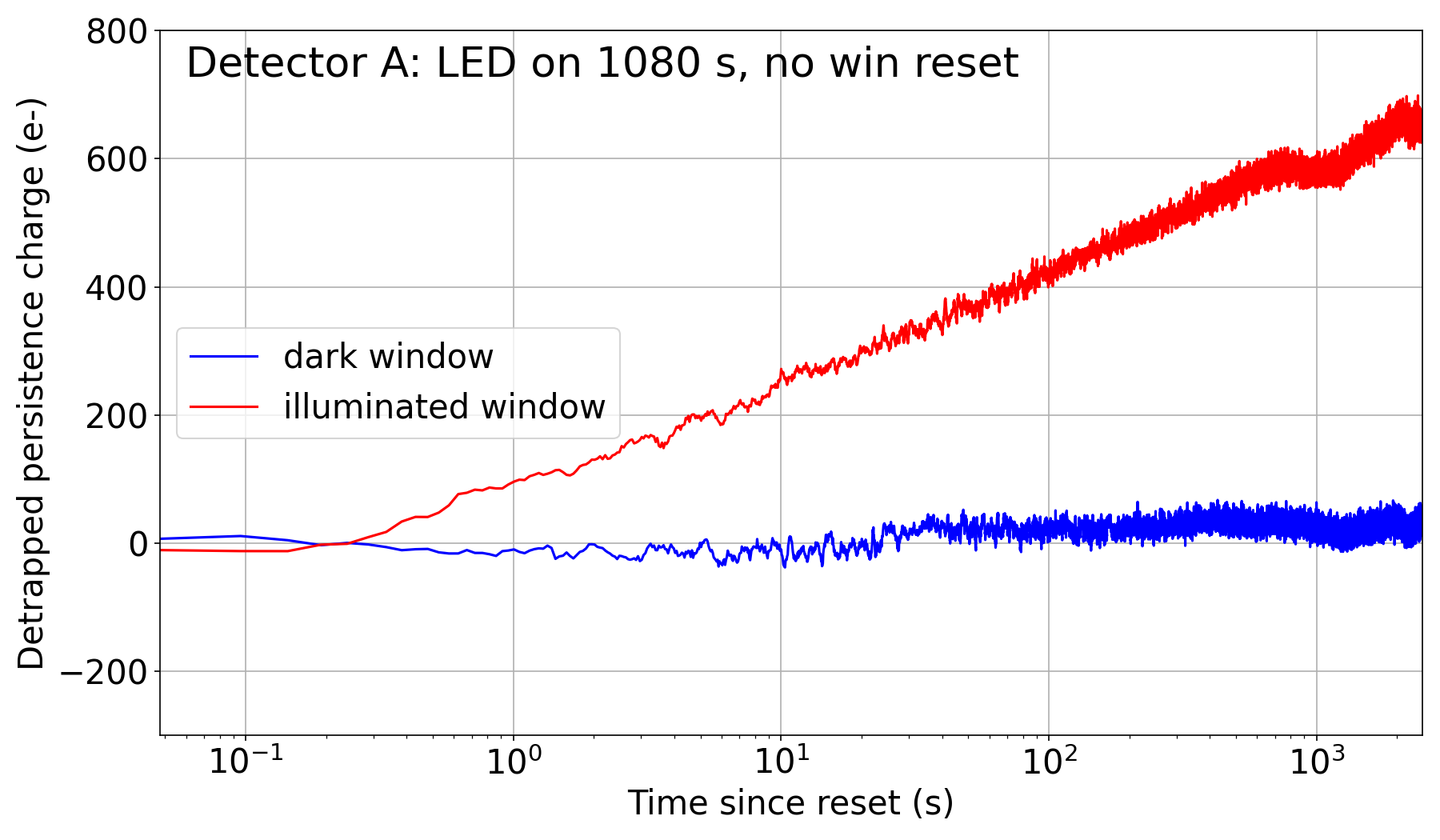}
        \includegraphics[width=0.8\linewidth]{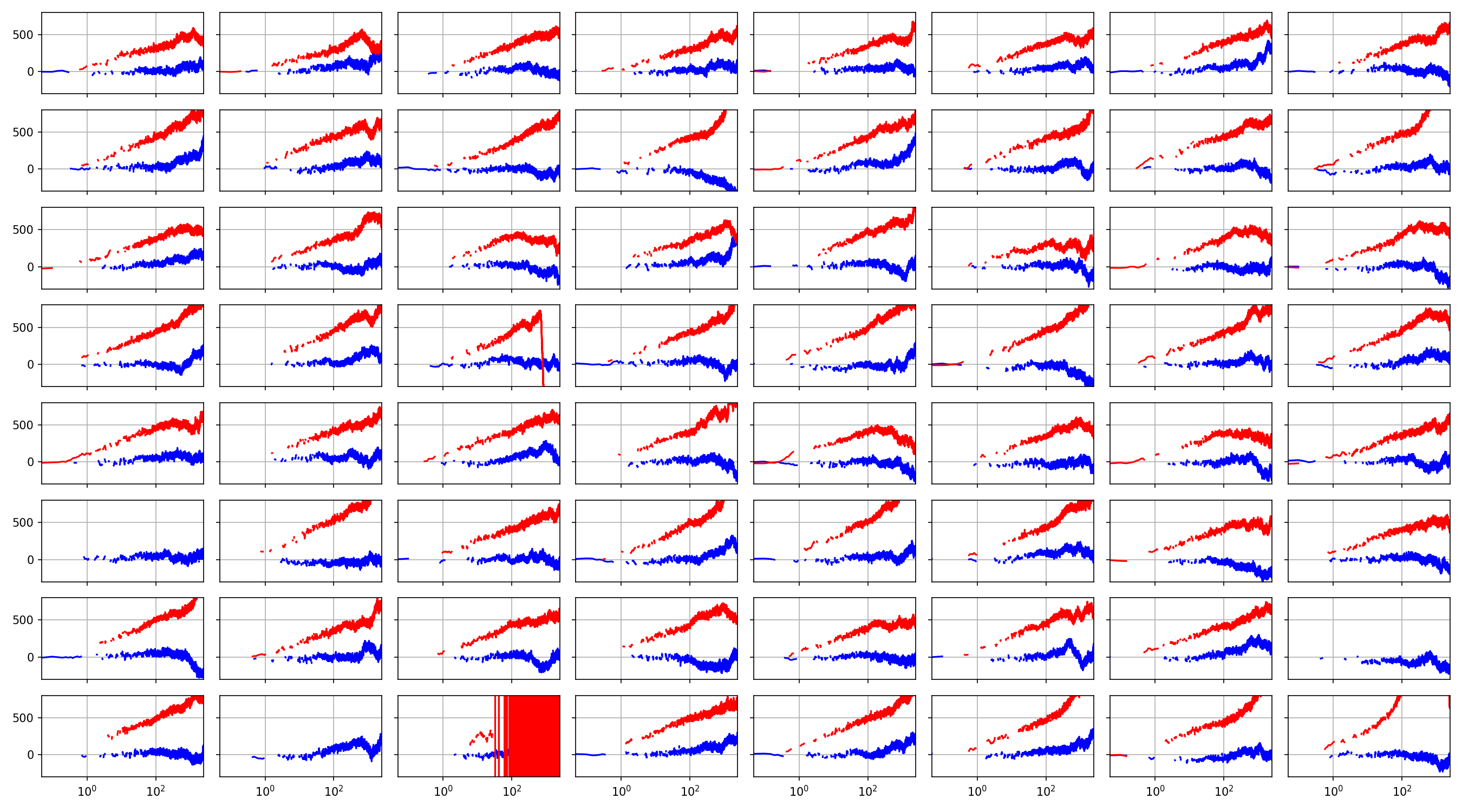}
    \end{center}  
    \caption{\label{fig:noReset}
    Example persistence measurement for Detector A during which the LED was activated for 1080\,s, and the windows were not held in reset. The lines show the differences in linearized UTR samples between the preceding and following darks for the illuminated window (red) and the dark window (blue). The top plot shows the mean curve for all functioning pixels, while the bottom plots shows individual curves for each pixel. The following bad pixels (zero-indexed row,col) from the bottom plot are omitted from the analysis: $(0,1), (0,2), (1,7), (2,0), (4,2)$. 
    }
\end{figure}

An example of this analysis is shown in Figure~\ref{fig:noReset}. In this experiment the windows are simply used to perform reads, and as expected there is a strong persistence signal; the post-illumination darks have approximately $\sim$600\,e- of excess charge compared to the initial darks at the end of the 36\,min exposure, after 1080\,s of illumination, while the reference dark window detects the same background before and after illumination in the same amount of time. Each time series is smoothed with a 10-sample boxcar to minimize the noise and to emphasize longer-term trends. The top plot shows the mean curve for all of the pixels, with some slow deviations from the trend beginning to occur after about 10\,min. The per-pixel plots in the lower portion of the figure show that there is in fact significant drift in both the dark and illuminated windows on long time scales. The fact that these drifts appear to be mostly uncorrelated from pixel to pixel suggests that the mechanism is not something simple like slowly changing thermal backgrounds. Note that the amount of detrapped charge that we detect in this experiment is broadly consistent with past results for comparable soak times\cite{tulloch2018}.

\begin{figure}[hbt]
    \begin{center}
        \includegraphics[width=0.8\linewidth]{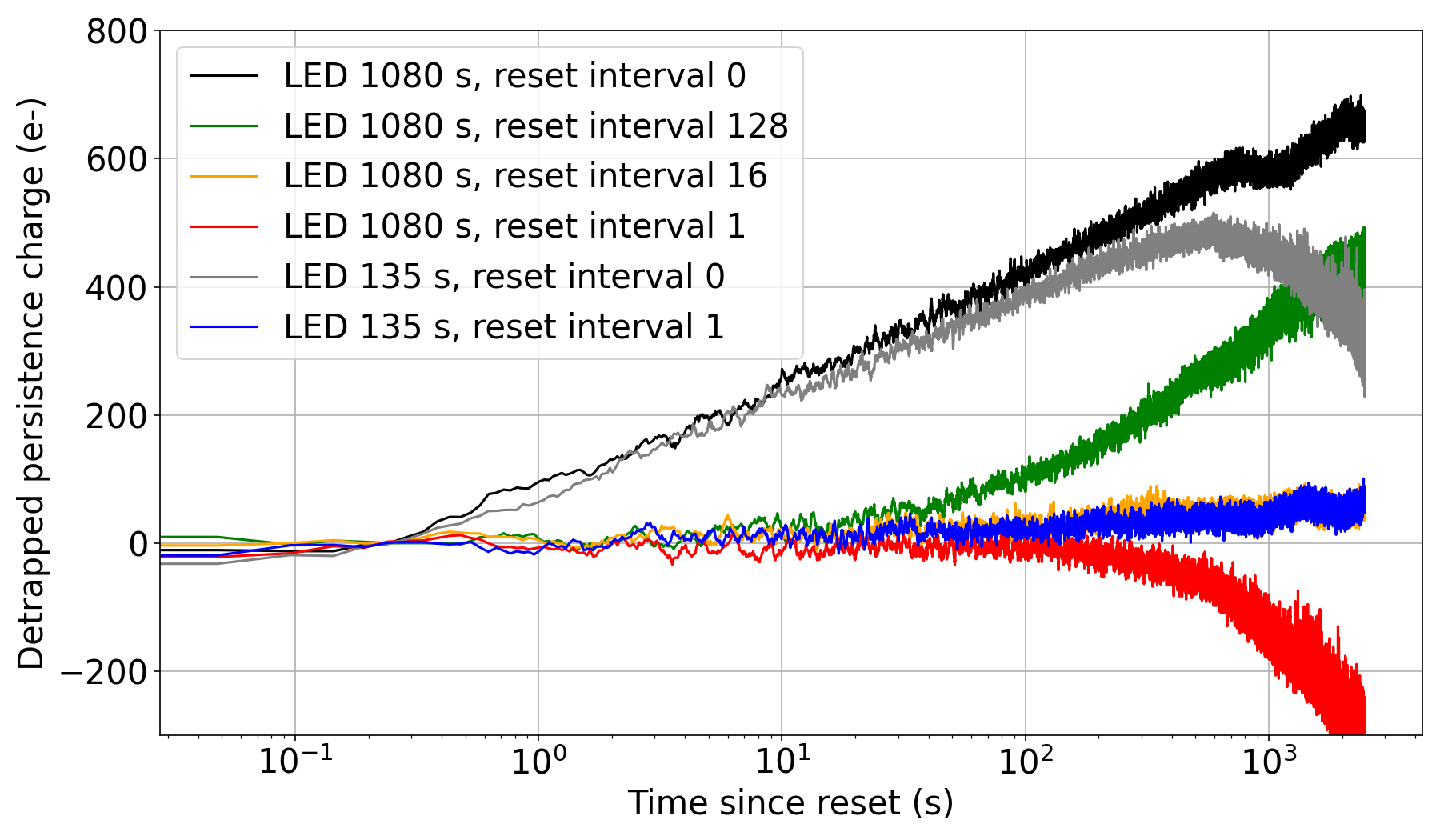}
    \end{center}  
    \caption{\label{fig:detAPersist}
    Summary of the persistence suppression experiments for Detector A. The effect of persistence is at a maximum when the windows are not periodically reset (black and grey curves). When resets are implemented at a rate that holds the signal at or below $\sim$20\% of the full-well depth (orange, red, and blue curves) persistence is not observed. At an intermediate reset rate that allows pixels to briefly saturate (green curve) persistence is approximately $\sim$half of the non-reset cases. The source of the long-timescale drifts at $t\gsim500$\,s may be related to the linearity correction.
    }
\end{figure}

The main results from our tests with Detector A are shown in Figure~\ref{fig:detAPersist}. Each curve shows the measured detrapped persistence charge averaged over all pixels in the illuminated window. As expected the two tests for which the windows were not reset exhibit the largest persistence, and the longer soak time (1080\,s, black curve) produces more of an effect than the shorter soak time (135\,s, grey curve). It is also clear that periodically resetting the window is highly effective at suppressing persistence: in the cases where the window is reset either every visit (blue and red curves), or every 16 visits (orange curve), the pixels never accumulate more than $\sim$20\% of the full-well depth in charge during illumination, and appear to be consistent with zero persistence. For the intermediate case of resetting the window every 128 visits (green curve) the pixels are saturated about 40\% of the time, and the detrapped charge is roughly half of what it is in the non-reset cases. 
However, the long-timescale ($\gsim$500\,s) behavior continues to exhibit the large excursions that were noted earlier for some of the curves (the downward-curving grey and red plots around 500\,s, and 200\,s respectively, and the ``kink'' in the black plot around 1000\,s), while the remaining curves exhibit something closer to the expected behavior (green, orange, and dark blue curves).

One potential explanation for this long-timescale behavior is that the ramp fits to the preceding darks are inaccurate and introduce growing errors when used to linearize the data from the following darks the further the pixel goes into the non-linear regime, although we lack well-sampled flatfield ramps (stacking many high-SNR observations) that could help us assess the quality of the individual dark pixel ramps that we are using here. Regardless, our measurements seem to provide consistent up to at least $\sim$200\,s.

\subsection{Peculiarities uncovered in Detector B experiments}

\begin{figure}[hbt]
    \begin{center}
        \includegraphics[width=0.8\linewidth]{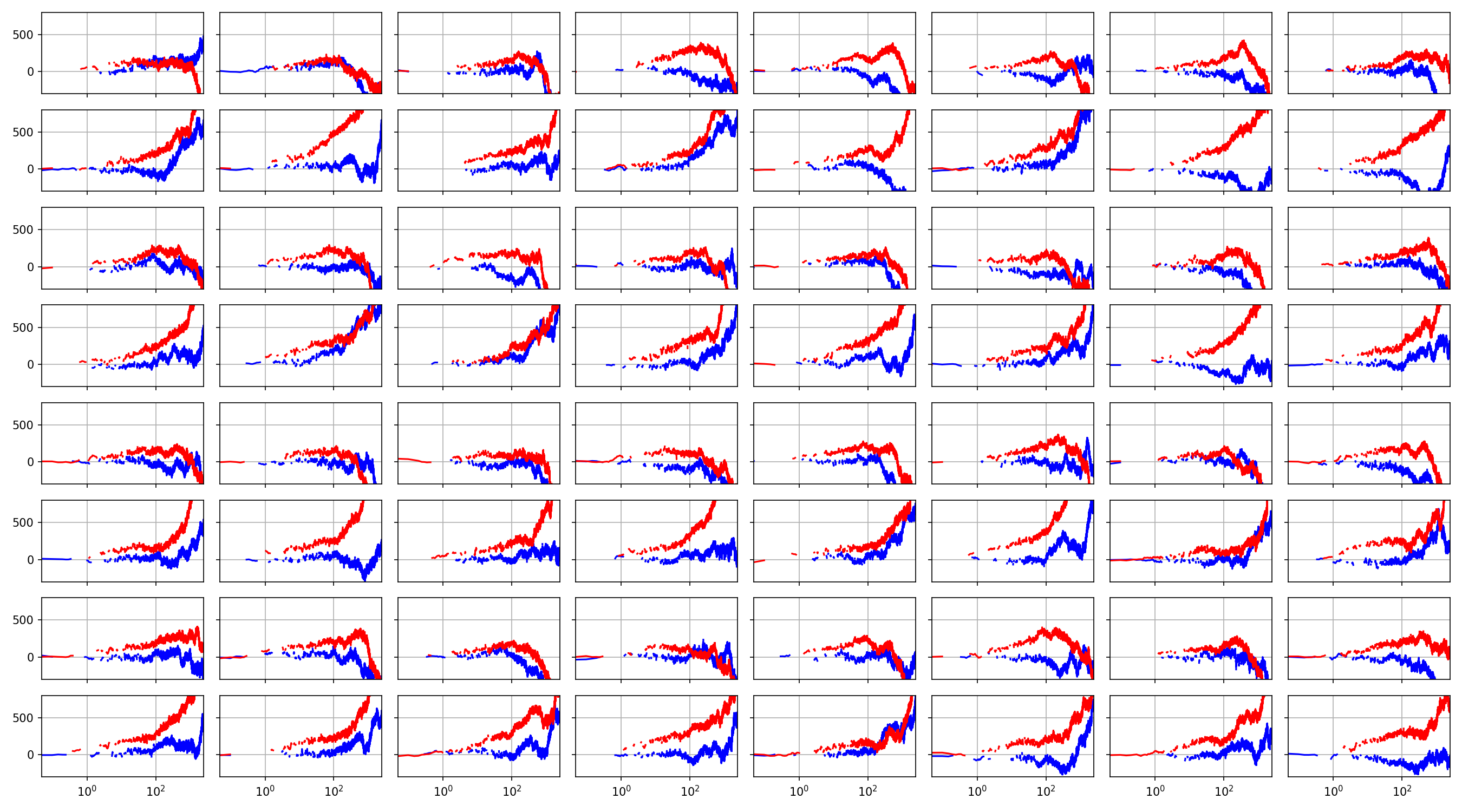}
    \end{center}  
    \caption{\label{fig:noResetB}
    Example persistence measurement for Detector B under the same conditions as Figure~\ref{fig:noReset}. The red curves are for the illuminated window, and blue curves are for the dark window. There is a much stronger long-timescale drift in the detrapped charge curves, and with a pattern that alternates by row not seen in Detector A.
    }
\end{figure}

Once Detector B was mounted in our test cryostat we repeated the test sequence and immediately noticed peculiar results. The equivalent curves of mean detrapped persistence charge exhibited less of an effect in the non-reset cases, and only a marginal improvement in the reset cases, as compared with Detector A. An example of the per-pixel persistence measurements without resetting the windows is show in Figure~\ref{fig:noResetB}.

\begin{figure}[hbt]
    \begin{center}
        \includegraphics[width=0.8\linewidth]{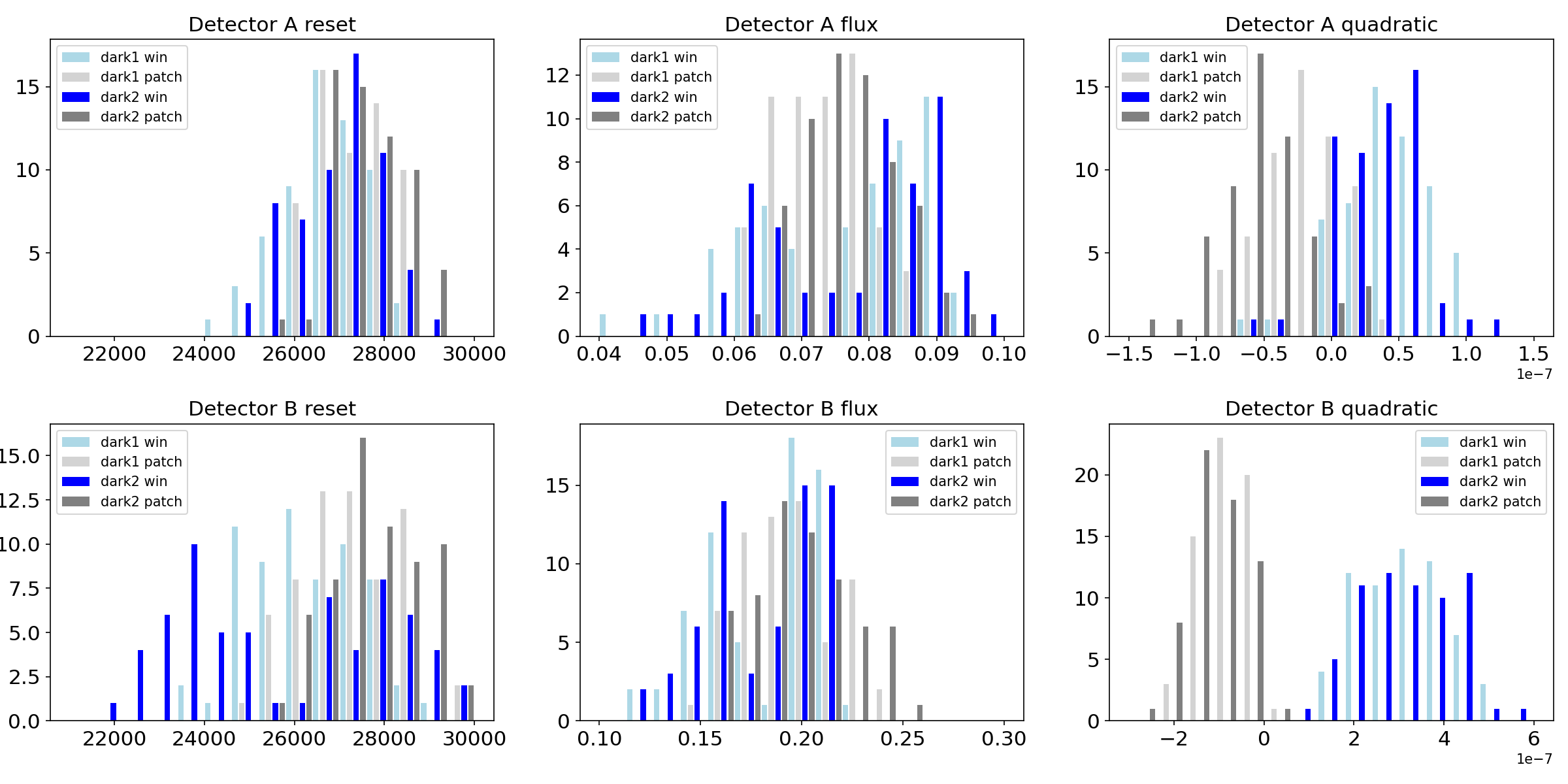}
    \end{center}  
    \caption{\label{fig:coeffHist}
    Histograms showing the spread in polynomial fit coefficients to window ramps, and also a patch of nearby science pixels with similar illumination, for both the preceding dark and the dark that follows illumination.
    }
\end{figure}

\begin{figure}[hbt]
    \begin{center}
        \includegraphics[width=0.8\linewidth]{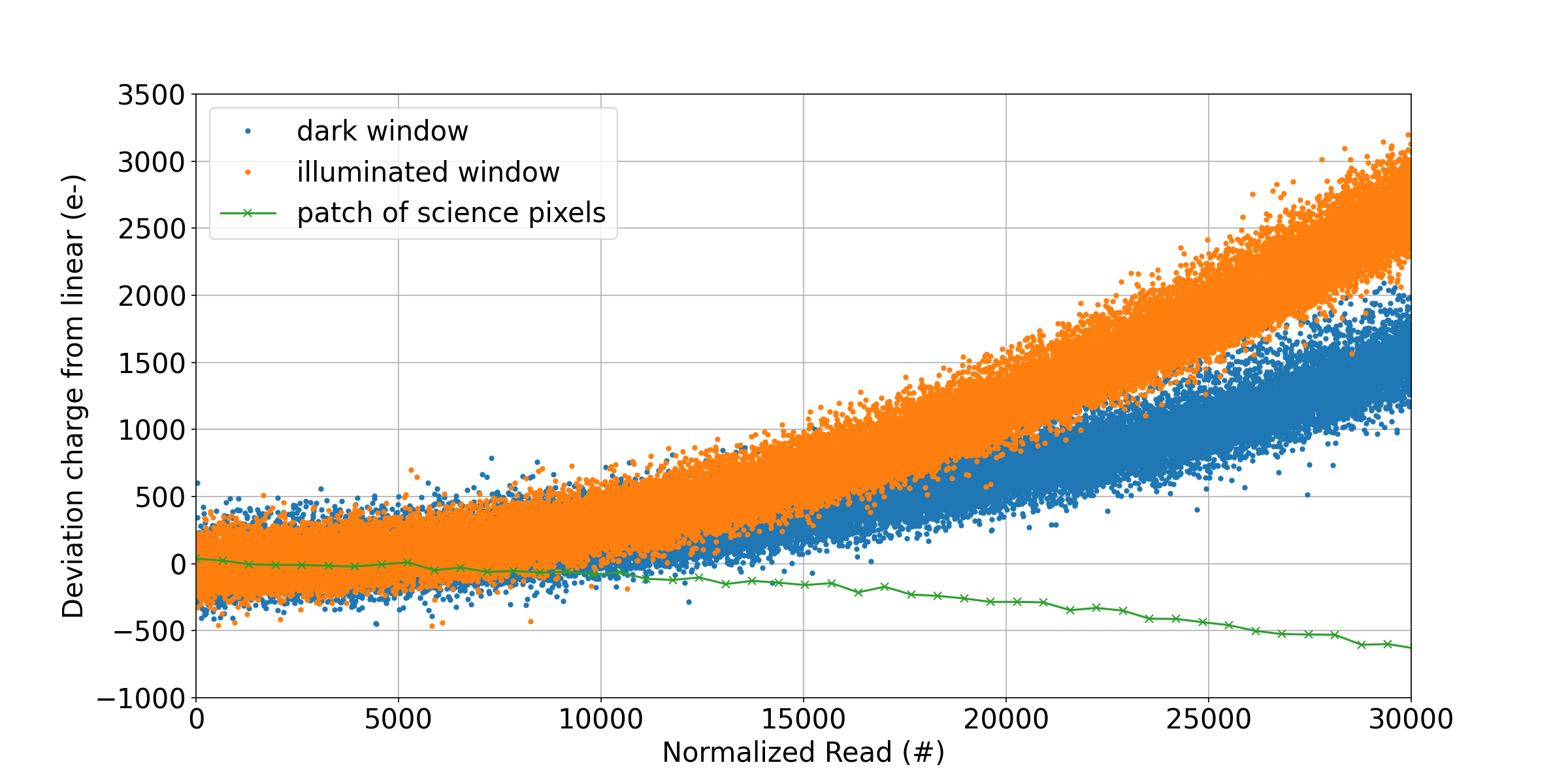}
    \end{center}  
    \caption{\label{fig:rampDiffs}
    Plot comparing flux-normalized ramp residuals once the linear components have been removed for window pixels (blue and orange) with a patch of nearby science pixels with similar illumination (green). These curves show positive deviations for the windows, and negative deviations for the science pixels.
    }
\end{figure}

\begin{figure}[hbt]
    \begin{center}
        \includegraphics[width=0.8\linewidth]{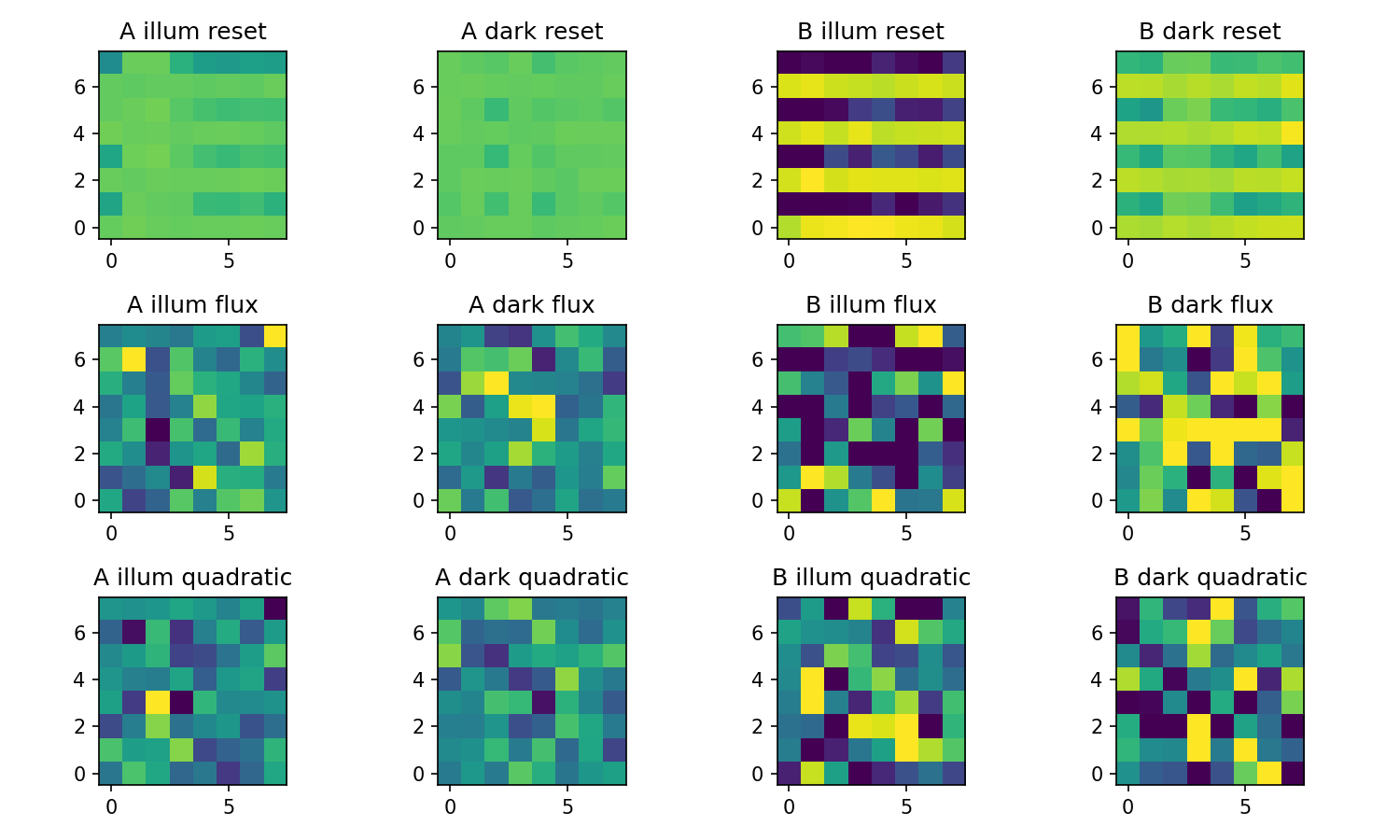}
    \end{center}  
    \caption{\label{fig:coeffDiffs}
    Maps showing the differences in fitted ramp coefficients for window pixels before and after illumination with the LED. The columns enumerate the detector and the window for which the measurement is made. Rows show maps of differences in reset level, flux, and quadratic terms. For this example the windows were reset during illumination, but the results are very similar when the windows are not being reset.
    }
\end{figure}

A striking feature of the data taken with Detector B is a pattern of long-timescale variations similar to that observed with Detector A, but now with very strong correlations in the signal along rows, and with a pattern that alternates by row. For example, the top row and every other row in Figure~\ref{fig:noResetB} shows both the illuminated and dark window persistence curves dipping down at large times (although the illuminated windows still tend to show greater charge, suggesting that there is also a persistence effect). The pattern is reversed for the even rows where the curves instead bend upward. We found similar trends whether or not the windows were being reset, though on short timescales there does seem to be a measurable reduction in persistence as seen in Detector A.

Examining the 2nd-order polynomial fits to each pixel ramp in the windows we identified other (potentially related) features. Figure~\ref{fig:coeffHist} shows histograms of the polynomial fit coefficients to the window pixel ramps, and also a patch of science pixels near the centre of the detector that received a comparable flux to the window-induced glow, produced by the ``ring'' of background emission noted in the left panel of Figure~\ref{fig:overview}. For Detector A (top row of Figure~\ref{fig:coeffHist}) all of the fit coefficients are comparable for data taken before and after illumination with the LED. However, for Detector B, there is a significant change whereby roughly half of the pixels in the illuminated window drop to lower values post-illumination (the bi-modal dark blue histogram at the bottom-left). Since the initial dark ramp fits do not encompass these lower reset levels, this effect could introduce linearity correction errors (because it is extrapolating the fit far from where it was originally measured).

Referring to the histograms of quadratic coefficients in Figure~\ref{fig:coeffHist} (rightmost column), it is also clear (particularly for Detector B) that they are oddly positive for the window pixels. In other words, unlike the more typical-looking ramp obtained through LED illumination in Figure~\ref{fig:ledWinRamp}, these background / self-heating dominated ramps in the dark bend {\em upwards} over time, rather than downwards, only for the window pixels. Another view of this phenomenon is shown in Figure~\ref{fig:rampDiffs}. A mean ramp is constructed for each collection of pixels (dark and illuminated window pixels, and the patch of science pixels), and a 2nd-order polynomial is fit. The curves are first normalized by flux (the $x$-axis is multiplied by the fitted slope) such that they would all follow the line $x=y$ if there were no non-linearities. The linear portion of the fit is then subtracted and the residuals are plotted. Note that at the right-hand side of this plot, the pixels have an integrated signal of $\sim$30000\,e- (roughly $\sim$30\% of the full-well depth). We verified that the slower full-frame reads of those same pixels show the same behavior (i.e., this is not a property of the window readout, rather the pixel itself). One interpretation might be that the window glow has a significant time-varying component (e.g., something like a thermal lag which causes it to grow over time). Smith and Hale (2012)\cite{smith2012} performed tests with fast windowed reads, and would visit small $8\times8$ windows ``for long enough to establish thermal equilibrium'' before increasing the window sizes, which suggests that such a lag may be an important effect, although they do not report on the timescales involved.

To explore the spatial distribution of these patterns, Figure~\ref{fig:coeffDiffs} shows maps of differences in the window fit coefficients before and after illumination. The flux (slope) and quadratic terms (bottom two rows) do not exhibit any strong patterns. However, the reset levels do exhibit both row- and column-correlated patterns. The strongest pattern is that shown for the reset level of the illuminated window in Detector B which is clearly the same feature as the bi-modal change in reset levels post-illumination noted in Figure~\ref{fig:coeffHist}. To re-state what we are observing: there is an apparently persistent change in pixel reset levels caused by {\em the combination of bright illumination and frequent window visits}, and it seems to occur whether or not the window is also being reset. Eventually the reset levels relax to their original values (as observed when a new set of observations are performed the next day); in a future experiment it would be valuable to reset the detector and conduct new darks periodically after the completion of the main illumination experiment to track its time evolution. Another useful variation would be to conduct these follow-up darks without using windows to observe how long it takes science pixels to revert to their original behavior.

%
%
\section{Summary and future work}

We conducted a series of experiments using two different 5-$\mu$m cutoff HAWAII-2RG sensors to establish the utility of performing fast localized resets at locations of bright sources, in parallel with long integrations of the full sensor (for science), as a mitigation for persistence. This general idea has been proposed for upcoming instruments such as the InfraRed Imaging Spectrograph (IRIS) for the Thirty Meter Telescope. Updates to our test system over an earlier experiment include improved temperature stability, use of a compact light source (leaving some portions of the detector in the dark as a reference), and cosmetically superior chips. For the first detector tested we obtained highly promising results: we were able to induce and measure persistence comparable to past experiments, and then showed that persistence could be removed by resetting the windows fast enough that pixels never saturate. Even operating the windows at lower rates that allow the pixels to saturate for short periods offered a benefit. However, experiments with the second detector provided mixed results, and revealed some odd behaviors. First, long-timescale drifts in our persistence measurements (which consist of differences between time-series of non-destructive reads performed in the dark before and after illumination) made it particularly difficult to quantify our results in the case of the second detector. We noted a strong alternating correlation in this pattern along rows. Further investigation showed that our second detector experiences a strong bi-modal shift in window pixel reset levels which seems to be induced by the combination of bright illumination with frequent window visits (whether they are reset or simply read each time). Another strange feature noted for both of the detectors tested is that pixel ramps for the windows have atypical shapes for the dark observations. In these cases the window pixels observe a thermal background, as well as glow induced by the window visits (with the latter being $\sim$3--6 times brighter than the former). Second-order polynomial fits to these ramps yield {\em positive} quadratic terms (i.e., they bend ``up'' with respect to the linear portion of the curve). However, when these same pixels are illuminated by the LED they exhibit more typical ramps.
A potential explanation is that there is a significant thermal lag in the window glow which causes it to grow over an initial period; this effect is sub-dominant in the case of LED illumination which would explain the different ramp shapes.

The general results from our tests show promise for the use of fast window resets to minimize persistence, but it is clear that there is a wide variance in the behavior of particular detectors. While the long-timescale drifts in our dark measurements may only reflect problems with our test system or ramp linearization strategy, the apparently window-induced reset level changes caused in regions of the detector that are illuminated is a potential cause for concern. It would be useful to conduct future experiments with repeated dark exposures (initiated after a reset) following illumination to determine whether these effects are truly localized to the windows, and if they are, typical timescales for them to relax to their original values.

\acknowledgments

We would like to thank co-op students Owen Hubner and Mustafa Wasif who helped with modifications to the Generation IV ARC controller software that enabled the hybrid full-frame / window readout mode that we used. This project was partially supported by a New Beginnings grant from National Research Council Canada.

\bibliography{refs} 
\bibliographystyle{spiebib} 

\end{document}